\documentclass[doublecol]{epl2} 
\def\be{\begin{equation}}
\def\ee{\end{equation}}

\def\bc{\begin{center}} 
\def\ec{\end{center}}
\def\bea{\begin{eqnarray}}
\def\eea{\end{eqnarray}}

\newcommand{\Avg}[1]{\left\langle{#1}\right\rangle}
 \usepackage{dcolumn}
\usepackage{bm}
\usepackage{graphicx}
\usepackage{color}
\usepackage{subfigure}
\usepackage{hyperref}
\usepackage{latexsym}
\usepackage{amsthm}
\usepackage{amssymb}

\DeclareGraphicsExtensions{.jpg,.pdf, .mps, .png, .eps, .ps, .EPS,.gif}

\title{Functional Multiplex PageRank}
\shorttitle{Functional Multiplex PageRank} 

\author{Jacopo Iacovacci$^1$, Christoph Rahmede$^2$, Alex Arenas$^3$ and Ginestra Bianconi$^1$}
\shortauthor{J. Iacovacci, C. Rahmede, A. Arenas and G. Bianconi}
\institute{$^1$School of Mathematical Sciences, Queen Mary University of London, London E1 4NS, United Kingdom\\
$^2$Rome International Centre for  Material Science Superstripes RICMASS, 
00185 Roma, Italy\\
$^3$Departament d'Enginyeria Inform\'atica i Matem\'atiques, Universitat Rovira i
Virgili, 43007 Tarragona, Spain}


\abstract{Recently it has been recognized that many complex  social, technological and biological networks have a  multilayer nature and can be described by multiplex networks. Multiplex networks are formed by a set of nodes connected by links having different connotations forming the different layers of the multiplex. Characterizing the centrality of the nodes in a multiplex network is a challenging task since the centrality of the node naturally depends on the importance associated to links of a certain type. Here we propose to assign to each node of a multiplex network a centrality called Functional Multiplex PageRank that is a function of the weights given to every different pattern  of connections (multilinks) existent in the multiplex network  between any two nodes. Since multilinks distinguish all the possible ways in which the links in different layers can overlap, the Functional Multiplex PageRank can describe important non-linear effects when  large relevance or small relevance  is assigned to multilinks  with  overlap. Here we apply the Functional Page Rank to the multiplex airport networks,  to the neuronal network of the nematode C. elegans, and to social collaboration and citation networks between scientists. This analysis reveals important differences existing between the most central nodes of these networks, and the correlations between their so called {\em pattern to success.}}
\begin{document}

\maketitle

\section{Introduction}

Many complex interacting systems are formed by nodes related by different types of interactions forming multiplex networks \cite{PhysReports,Kivela,Goh_review,Perspective}.
Examples of multiplex networks are ubiquitous, from social \cite{Mucha,Thurner,Menichetti,Jacopo} to transportation  \cite{Cardillo,Alex_rw,Latora} and biological networks \cite{Bullmore,Latora}.
For example scientific authors form at the same time collaboration networks and citation networks in which they cite each other \cite{Menichetti,Jacopo}, the airport network is formed by airports connected by flights operated by different airline companies \cite{Cardillo},  in the brain neurons are simultaneously connected by chemical and electrical types of connections \cite{Latora,DeDomenico1,datacelegans}.
A multiplex network is therefore constituted  by a set of $N$ nodes interacting through $M$ layers which are networks formed by links having the same connotation.
In recent years we have gained a significant understanding of the  interplay between the structure and the dynamics of multiplex networks \cite{diffusion,Boguna_epidemics,Havlin1,Doro_multiplex,BD1,game,Radicchi} and relevant insights regarding the level of  information encoded in  their correlated structure  \cite{Mucha,Menichetti,Jacopo,Latora,PRE,Goh_correlations,growing}.
In this context, given the increasing number of multiplex network datasets, establishing the centrality of the nodes in multiplex networks has become a problem
of major interest. 
Until now, several multiplex centrality measures have been proposed \cite{MPageRank,Jacopo,Romance,Versatility,Maths,DeDomenico2,Temporal}  which aim at going beyond the definition of centrality in single networks \cite{Page,Newman,Santo_PR}. Examples of multiplex centrality measure  include the  Versatility of the nodes \cite{Versatility}, the Multiplex PageRank \cite{MPageRank,Jacopo}, and the Eigenvector multiplex centrality\cite{Romance}. The Versatility \cite{Versatility} emphasizes the relevance of nodes connected in many different layers and it  applies to multiplex networks where corresponding nodes in different layers are connected by interlinks. The Multiplex PageRank \cite{MPageRank,Jacopo} exploits the correlations existing between the degree of the nodes in different layers through the use of a biased random walk.  The Eigenvector multiplex centrality \cite{Romance} instead assumes that the centrality of a node with respect to one layer is influenced by its centrality in other layers weighted by  a matrix of influences that one layer has on the other layer. Both the Multiplex PageRank and the Eigenvector multiplex centrality do not make explicit use of the interlinks. 
The main challenge when defining a centrality of the nodes in a multiplex network without interlinks is that the centrality depends on the relevance associated to the different types of possible interactions that can exist between the nodes.
Recently, the complete set of possible interactions existing between any two nodes of a multiplex network has been fully characterized using the {\em multilinks} \cite{PRE,Menichetti}. The multilinks specify all the layers in which any  pair of two nodes is  connected.
Here we propose a new centrality measure called Functional Multiplex PageRank which assigns to each node a centrality depending on the influence assigned to each type of multilink.
In this way it is possible to go beyond the modelling of the influence of each single layer because the influence of multilinks can capture important non-linear effects due to the overlap  links. For example in a duplex network it allows us to weight a connection existing in both layers much more or much less than the sum of the weights attributed to connections  which are exclusively present in one  of the layers.
The Functional Multiplex PageRank associated to each node is a function also called the {\em pattern to success} of the  node.
The Functional Multiplex PageRank allows the comparison of the pattern to success corresponding  of different nodes and important insights can be gained by studying their correlations.
Finally starting from the calculation of the Functional Mutliplex PageRank it  is possible to extract an Absolute Multiplex PageRank which is able to provide a unique ranking for all the nodes of the multiplex network.

\section{Multiplex networks and multilinks}
A  multiplex  network  $\vec{G}=(G_1,G_2,\ldots, G_M)$ is formed by a set of  $N$ nodes $V$ and $M$ layers (networks) $G_{\alpha}=(V,E_{\alpha})$ with $\alpha=1,2,\ldots, M$ and $E_{\alpha}$ indicating the set of links in layer $\alpha$.  It is to be noted here that in  this definition of a multiplex network we do not include explicitly the interlinks and therefore the multiplex networks we are considering are equivalent to  colored networks where links of different colors form the different layers.

The multiplex network that we consider in this paper  can be fully represented using $M$ adjacency matrices ${\bf a}^{[\alpha]}$ with $\alpha=1,2,\ldots M$ indicating the interactions occurring in each layer $\alpha$.
We assume that the networks forming the different layers are directed and unweighted and we adopt the convention that the adjacency matrix element $a_{ij}^{[\alpha]}=1$ if and only if node $j$ points to node $i$ in layer $\alpha$, otherwise $a_{ij}^{[\alpha]}=0$.

In  multiplex networks two generic nodes can be connected in more than two layers. 
In this case we say that there is a link overlap in the multiplex network.
The significance of the  overlap of the links between any two layers $\alpha,\beta$ can be evaluated by means of the total  overlap between any two layers \cite{PRE}. The total overlap $O^{[\alpha,\beta]}$ between layer $\alpha$ and layer $\beta$  is given  by the total number of pairs of nodes connected both in layer $\alpha$ and layer $\beta$, i.e. 
\bea
O^{[\alpha,\beta]}=\sum_{i, j}a_{ij}^{[\alpha]}a_{ij}^{[\beta]}.
\eea
For a general multiplex network with $M>2$ there are multiple ways in which the links can  overlap across different layers.
A way to fully characterize multiplex networks with link overlap is to use the recently introduced concept of multilinks \cite{PRE,Menichetti}. In fact, in a   multiplex network it is convenient to specify for each pair of nodes $i$ and $j$ all the layers in which they are connected. This is simply achieved by indicating which type of {\em multilink} $\vec{m}$  connects the two generic nodes.
We first define the vector $\vec{m}=(m_1,m_2,\ldots, m_M)$ of generic elements $m_{\alpha}=0,1$ with $\alpha=1,2,\ldots M$.
We say that a generic   pair of  nodes $(i,j)$ is connected by a multilink $\vec{m}=\vec{m}^{ij}$ if   and only if $\vec{m}^{ij}=(a_{ij}^{[1]},a_{ij}^{[2]},\ldots, a_{ij}^{[M]})$. In particular, two nodes that are not connected in any layer are connected by the {\it  trivial multilink} $\vec{m}=\vec{0}$.

The multi-adjacency matrices $\bf A^{\vec{m}}$  define the neighbors of a node that are connected with a multilink of type $\vec{m}$. Specifically the multi-adjacency matrix elements take the value one ($A_{ij}^{\vec{m}}=1$) if the pair of nodes $(i,j)$ is connected by a multilink $\vec{m}$, while  otherwise $A_{ij}^{\vec{m}}=0$. Therefore the multi-adjacency  matrix elements $A_{ij}^{\vec{m}}$ can be expressed in terms of the adjacency matrices ${\bf a}^{[\alpha]}$ as  
\bea
A^{\vec{m}}_{ij}=\prod_{\alpha=1}^M \left[m_{\alpha}a_{ij}^{[\alpha]}+(1-m_{\alpha})(1-a_{ij}^{[\alpha]})\right].
\eea
Since every pair of nodes $(i,j)$ can be connected by a unique multilink, the multiadjacency matrices satisfy 
\bea
\sum_{\vec{m}}A_{ij}^{\vec{m}}=1.
\eea
Moreover we have $A_{ij}^{\vec{m}}=1$ for the unique type of multilink $\vec{m}=\vec{m}^{ij}$, i.e.
$A_{ij}^{\vec{m}^{ij}}=1.$

Take for example the case of a duplex network formed by two layers $\alpha=1,2$. The multiadjacency matrices corresponding to non trivial multilinks $\vec{m}\neq \vec{0}$ are given by  
\bea
A_{ij}^{(1,0)}&=&a_{ij}^{[1]}(1-a_{ij}^{[2]})\nonumber \\
A_{ij}^{(0,1)}&=&(1-a_{ij}^{[1]})a_{ij}^{[2]}\nonumber \\
A_{ij}^{(1,1)}&=&a_{ij}^{[1]}a_{ij}^{[2]}.
\eea
Assume that we consider the airport duplex networks formed by two layers corresponding to the flight connections of two different airline companies. In this case $A_{ij}^{(1,0)}=1$ if the airport $i$  can be reached from airport $j$ by a direct flight connection of the first company but not by a direct flight connection of the second company, $A_{ij}^{(0,1)}=1$ if the airport $i$  can be reached from airport $j$ by a direct flight connection of the second company but not by a direct flight connection of the first company, and finally $A_{ij}^{(1,1)}=1$ if airport $i$ can be reached from airport $j$ by direct flight connections of both airline companies.

In general the multilinks can characterize all the different ways in which a pair of nodes of the multiplex network can be connected across the $M$ layers. Nevertheless the number of different types of multilinks increases exponentially with the number of layers $M$, as it is $2^{M}$, and for large number of layers the distinction between different multilinks can be problematic.
A way to go around this problem is to classify all the multilinks according to the {\em multiplicity of link overlap} $\nu$ given by 
\bea
\nu(\vec{m})=\sum_{\alpha=1}^M m_{\alpha}.
\eea 
Therefore if node $i$ and node $j$ are linked by a multilink $\vec{m}$, the multiplicity of the overlap  $\nu(\vec{m})$ indicates in how many layers there is a connection between the two nodes.
If we only distinguish between multilinks with different multiplicity of the overlap, we can drastically reduce the complexity of the analysis because the range of variability of $\nu(\vec{m})$ is given only by the set $\{1,2,\ldots, M\}$.

\section{Functional Multiplex PageRank}

In this section we will define a  Functional Multiplex PageRank  $X_i$ of node $i$. This centrality measure depends functionally on a set of parameters $\bf z$ and  as a function of the parameters ${\bf z}$ can be reduced to:
\begin{itemize}
\item PageRank on each separate layer; 
\item PageRank on the aggregated network;
\item PageRank on the network formed by the links ($i,j$) present at the same time  in every layer $\alpha=1,2,\dots M$.
\end{itemize} 
The Functional Multiplex PageRank $X_i$ of node $i$ depends on the  tensor  ${\bf z}$ with elements $z^{\vec{m}}\geq 0$ defined for every type of multilink $\vec{m}$. It describes the steady state of a random walker that hops from a node $j$ to a neighbor node $i$ with probability $\tilde{\alpha}$ if this is possible, and otherwise  performs random jumps to a random connected node of the multiplex network.
When the random walker hops to a random neighbor it  follows each multilink $\vec{m}\neq \vec{0}$ with a probability proportional to $z^{\vec{m}}$. Therefore we have that the Functional Multiplex PageRank $X_i(\bf{z})$ of node $i$ is given by  
\bea
X_i({\bf z})=\tilde{\alpha} \sum_{j=1}^N A_{ij}^{\vec{m}^{ij}}z^{\vec{m}^{ij}}\frac{1}{\kappa_j}X_j+\beta v_i,
\label{Xz}
\eea
where $z^{\vec{0}}=0$ and where
\bea
\kappa_j &= &\sum_{i=1}^N A_{ij}^{\vec{m}^{ij}}z^{\vec{m}^{ij}}+\delta_{0,\sum_{i=1}^N A_{ij}^{\vec{m}^{ij}}z^{\vec{m}^{ij}},}\nonumber\\
\beta &= &\frac{1}{N}\sum_{j=1}^N \left[(1-\tilde{\alpha})(1-\delta_{0,\sum_{i=1}^N A_{ij}^{\vec{m}^{ij}}z^{\vec{m}^{ij}}}),\right.\nonumber \\
&&+ \left.\delta_{0,\sum_{i=1}^N A_{ij}^{\vec{m}^{ij}}z^{\vec{m}^{ij}}}\right]X_j,\nonumber \\
v_i &=& \theta\left(\sum_{j=1}^N A_{ij}^{\vec{m}^{ij}}z^{\vec{m}^{ij}}+\sum_{j=1}^N A_{ji}^{\vec{m}^{ji}}z^{\vec{m}^{ij}}\right).
\eea
Here $\delta_{x,y}$ indicates the Kronecker delta and $\theta(x)$ indicates the Heaviside step function.
Using the definition of the Functional Multiplex PageRank given by Eq. $(\ref{Xz})$, by making an opportune choice of the influences ${\bf z}$ we can  recover the desired limiting cases:
\begin{itemize}
\item for $z^{\vec{m}}=z^{\star}>0$ for every $\vec{m}$ with  $m_{\alpha}=1$, and $z^{\vec{m}}=0$ for every $\vec{m}$ with $m_{\alpha}=0$, the Functional Multiplex PageRank reduces to the PageRank in the layer $\alpha$;
\item
for $z^{\vec{m}}=z^{\star}>0$ for every $\vec{m}\neq \vec{0}$, the  Functional Multiplex PageRank reduces to the PageRank in the aggregated network;
\item 
for $z^{\vec{m}}=0$ for every $\vec{m}\neq \vec{1}$, and $z^{\vec{1}}=z^{\star}>0$, the  Functional Multiplex PageRank reduces to the PageRank in the network where each link $(i,j)$ is present in every layer, i.e. $(i,j)$ are connected by a multilink $\vec{1}$.
\end{itemize}
{For every node $i$ the Functional Multiplex Centrality $X_i({\bf z})$ is  a function depending on the values of the influences ${\bf z}$ associated to its multilinks.   We  call this function the  {\em pattern to success} of node $i$.}

In general the Functional Multiplex PageRank allows the association of   different influence $z^{\vec{m}}$ to each type of multilink and can therefore be used to capture the specific role of overlapping links.
Consider for example the case of a duplex network. The Functional Multiplex PageRank  can  describe  non-linear effects due to the overlap between the links when the weight associated to a multilink ${(1,1)}$ is not  equal to  the sum of the weights associated to the multilinks $(1,0)$ or $(0,1)$, i.e. when  
\bea
z^{(1,1)}\neq z^{(1,0)}+z^{(0,1)}.
\eea

From the definition of the Functional Multiplex PageRank one observes that the ranking ${\bf X}({\bf z})$ is invariant under the transformation 
\bea
{\bf z}=\gamma{\bf z}
\eea
for $\gamma>0$.
Therefore by considering ${\bf z}$ as a vector in a $2^{M}-1$ dimensional space,  with elements $z^{\vec{m}}$ for every  $\vec{m}\neq \vec{0}$, the Functional Multiplex PageRank only depends on the direction of this vector and not on its normalization. Therefore the general definition of the Functional Multiplex PageRank depends on $2^{M}-2$ independent parameters.

We make here a general remark: although the above definition is very general, the definition of the Functional Multiplex PageRank for more than two layers might yield a centrality depending on too many parameters ($2^{M}-2$) and can be difficult to handle. Therefore in the following we will describe first the case of a duplex network that can be treated in full generality using two parameters, and subsequently we will cover the case of a general multiplex network $M$ where the  parameters $z^{\vec{m}}$ are taken to be dependent on  a single external parameter $q$ and on the multiplicity of the overlap $\nu(\vec{m})$.

{ Note that here   and in the following we have  considered  unweighted multiplex networks. Consequently the  multiadjacency matrices have elements $A_{ij}^{\vec{m}}=0,1$. However it is possible to consider a weighted version of the Functional Multiplex PageRank just by  using in Eq. (\ref{Xz}) a set of weighted multiadjacency matrices. The weighted multiadjacency matrices  associate a weight to each  multilink. This weight  could be equal, for instance, to the average  weight of the  links forming the multilink, or, alternatively  to their maximum or minimum weight.}

\section{Absolute Functional Multiplex PageRank}

It is often the case that  one desires a global  ranking of the nodes. Here we propose to take as the ranking of node $i$ the maximum of the Functional  Multiplex PageRank over all the space in which the vector ${\bf z}$ varies. To this end we define the Absolute Multiplex PageRank $X_i^{\star}$ of  node $i$ given by  
\bea
X_i^{\star}=\max_{\bf z} X_i(\bf z).
\eea
This constitutes only a possible choice of establishing an Absolute Rank from the Functional Multiplex PageRank.
Another interesting possibility is to consider the absolute ranking induced by the average of the Functional Multiplex PageRank, i.e.
\bea
\hat{X}_i=\Avg{X_i({\bf z})}_{\bf z}.
\eea 
In the following we will refer exclusively to the Absolute Multiplex PageRank $X_i^{\star}$.
 
 \section{Understanding Functional Multiplex Pagerank, a geometric interpretation}
 
 In this section we will study Functional Multiplex PageRank on different duplex networks showing the rich information that can be gained about its nodes.
 In all the cases discussed below we will span all the region of variability of ${\bf z}=(z^{(1,0)},z^{(0,1)}, z^{(1,1)})$ and we will always take $\tilde{\alpha}=0.85$ which is the usual value that  is considered for the PageRank on single layers.
 As  has been discussed in the previous section, the Functional Multiplex PageRank only depends on the direction of ${\bf z}$ interpreted as a 3-dimensional vector in $\mathbb{R}^3$.
Therefore in a duplex network, changing only  the direction of the vector ${\bf z}$ within the 3-dimensional region where  all the components of ${\bf z}$ are either positive or null,  is sufficient to span all cases. Therefore the different directions of ${\bf z}$ can be parametrized just by using two parameters.
Here we express  ${\bf z}=(z^{(1,0)},z^{(0,1)}, z^{(1,1)})$ in spherical coordinates as 
\bea
z^{(1,0)}&=&\sin\theta \cos \phi,\nonumber \\
z^{(0,1)}&=&\sin\theta \sin \phi, \nonumber \\
z^{(1,1)}&=&\cos \theta,
\eea 
with $\theta, \phi \in [0,\pi/2]$.

As a first duplex network on which to apply the functional PageRank we consider the airport duplex network formed by the flight connections of Lufthansa (layer 1) and British Airways (layer 2) constructed by using the data of Ref. \cite{Cardillo}.

The angle $\phi$ modulates the influence of the multilinks $(1,0)$ (exclusively  Lufthansa flight connections) with respect to multilinks $(0,1)$ (exclusively British Airways flight connections).
For $\phi=0, \theta=\pi/2$ or  $(\phi=0^o, \theta=90^o)$ the influence of exclusively Lufthansa connections is maximized, for $\phi=\pi/2, \theta=\pi/2$ (or $\phi=90^o, \theta=90^o$) the influence of exclusively British Airways is maximized.
The angle $\theta$ measures the influence of multilinks $(1,1)$ corresponding to flight connections existing in both airline companies with respect to the other two types of multilinks corresponding to flight connections existing in a single airline company. For $\theta=0$ or ($0^o$) the influence of multilinks $(1,1)$ is maximized, while for $\theta=\pi/2$  (or $90^o$) it is minimized.

The Absolute Multiplex PageRank of this duplex network ranks its four top central airports according to the rank shown in Table $\ref{Table1}$. 
\begin{table}
\caption{Top ranked airports according to the Absolute  Multiplex PageRank, in a duplex airport network formed by Lufthansa and British Airways airlines. Here in order to find the absolute Functional Multiplex PageRank we evaluated the Functional Multiplex PageRank for angles $(\theta, \phi)$ chosen on a grid with spacing $\delta \theta=\delta\phi=\pi/80$.}
\label{Table1}
\begin{center}
\begin{tabular}{cc}
Rank  & Airport \\
 1 & Heathrow Airport (LHR)\\
 2 & Munich Airport (MUC) \\
 3 & Frankfurt Airport (FRA)\\
 4 & Gatwick Airport (LGW) \\ 
\end{tabular}
\end{center}
\end{table}

\begin{figure}
\begin{center}
\includegraphics[width=0.75\columnwidth]{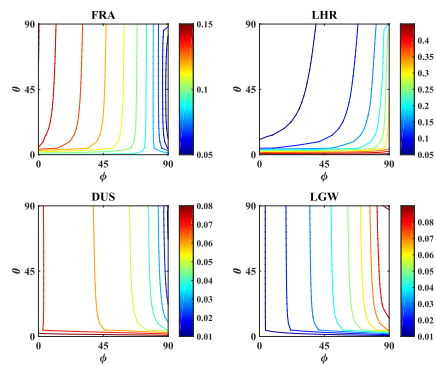}
\end{center}
\caption{(Color online) Functional Multiplex PageRank for important airports such as  Frankfurt Airport (FRA), Heathrow Airport (LHR),  D\"usseldorf Airport (DUS) and  Gatwick Airport (LGW).}
\label{figure1}
\end{figure}

We observe that the major airports  display a very different Functional Multiplex PageRank revealed by their distinct  {\it pattern to success}. In  Figure $\ref{figure1}$  we display the pattern of success of four exemplar hub airports. 
The Frankfurt airport (FRA) shows a pattern of success that establishes the airport as a central hub for Lufthansa. In fact its Functional Multiplex PageRank   displays a maximum  for smaller values of  $\phi$ and decreases as  $\theta$ decreases toward zero showing that the Frankfurt airport  takes most of its centrality from flight connections operated exclusively by Lufthansa.  On the  contrary, the D\"usseldorf airport (DUS) acquires significant centrality also by including  connections existing in both layers although it constitutes an important Lufthansa hub. Therefore it has a Functional Multiplex PageRank that is  increasing as  $\phi$ decreases, and also  increasing as $\theta$ approaches zero.
By calculating  the Functional Multiplex PageRank of Heathrow  Airport (LHR) and Gatwick Airport (LGW) one can see that they are both British Airways hub airports but Heathrow acquires important centrality also by including connections existing in both layers.

Therefore this result shows that the  links that determine the centrality of the nodes can be of different types for two different nodes of the multiplex network.
It is therefore interesting to measure the correlation between the Functional Multiplex PageRank (pattern to success) of two different nodes.
This  can be measured by considering the Functional Multiplex PageRank as a function of the angles $\phi,\theta$, i.e.  $X_i=X_i(\phi,\theta)$. By evaluating the Functional Multiplex PageRank on a given grid with points $(\phi_s,\theta_r)$, with  $r=1,2,\ldots N_{\phi}$ and $s=1,2\ldots N_{\theta}$  we can calculate the Pearson correlation $\rho$ between the Functional Multiplex PageRank of the generic nodes $i$ and $j$ as 
\bea
\rho=\frac{\overline{X_iX_j}-\overline {X_i}\ \overline{X_j}}{\sigma(X_i)\sigma(X_j)},
\eea
where $\overline{Y(\phi,\theta)}$
is given by 
\bea
\overline{Y(\phi,\theta)}=\frac{1}{N_{\phi}N_{\theta}}\sum_{r=1}^{N_{\phi}}\sum_{s=1}^{N_{\theta}}Y(\phi_s,\theta_r),
\eea
and $\sigma(Y)=\sqrt{\overline{Y^2}-\overline{Y}^2}$.
Note that $\rho\in [-1,1]$, where negative values $\rho<0$ indicate anticorrelations, while $\rho>0$ indicates positive correlations.  
 In Table $\ref{Table2}$ we report the correlations $\rho $ existing between the Functional Multiplex PageRanks shown in Figure $\ref{figure1}$ showing both positive (Heathrow/Gatwick, Frankfurt/D\"usseldorf but also Heathrow/D\"usseldorf) and negative values (Heathrow/Frankfurt,Gatwick/Frankfurt, Gatwick/D\"usseldorf).

\begin{table}
\caption{Correlation $\rho$ between the Functional Multiplex PageRank of the airports Heathrow (LHR), Frankfurt (FRA), Gatwick (LGW) and D\"usseldorf (DUS). The Pearson correlation is calculated starting form the Functional Multiplex PageRanks $X_i=X_i(\phi,\theta)$ calculated on a grid of $\delta \phi=\delta\theta=\pi/80$.}
\label{Table2}
\begin{center}
\begin{tabular}{ccccc}
$\rho$  &  LHR &   FRA &   LGW&  DUS\\
LHR &1& -0.797&0.484& 0.351\\
FRA&-0.797&1&-0.983& 0.275\\
LGW &0.484&-0.983&1&-0.729\\
DUS &0.351&0.2758&-0.729&1\\  
\end{tabular}
\end{center}
\end{table}

As a second example we  analyzed the multiplex connectome (brain network) of the neumatode C. elegans \cite{Latora,DeDomenico1,datacelegans}. This dataset includes all the connections existing between the 279 neurons of the animal. These connections can be chemical (synaptic connections forming the layer 1) or electrical (gap junctions forming the layer 2) \cite{Latora}. Both layers are undirected and unweighted.

\begin{table}[]
\centering
\caption{Top 10 ranked neurons according to the Absolute Multiplex PageRank of  the neuronal network of C. elegans where the first layer is formed by electric junctions and the second layer is formed by chemical synapses.
The Absolute Multiplex PageRank was performed using a grid $\delta\phi=\delta\theta=\pi/40$.}
\label{Tablece}
\begin{tabular}{cccc}
Rank & Neuron &Rank & Neuron\\
1    & AVAR   &6    & PVCR\\
2    & AVAL    &7    & AVDR \\
3    & AVBL   &8    & AVER \\
4    & AVBR   &9    & AVEL  \\
5    & PVCL  & 10   & DVA \\
\end{tabular}
\end{table}

\begin{figure}
\begin{center}
\includegraphics[width=0.75\columnwidth]{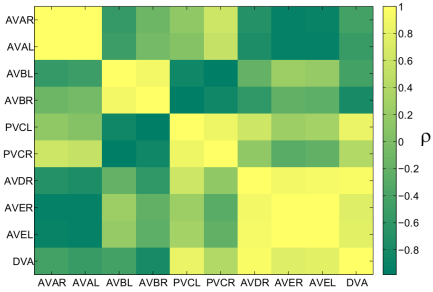}
\end{center}
\caption{(Color online) Correlation $\rho$ between the Functional Multiplex PageRank of the top ranked neurons in the  duplex brain network of the nematode C. elegans.}
\label{figurerho}
\end{figure}

  \begin{figure}
\begin{center}
\includegraphics[width=0.8\columnwidth]{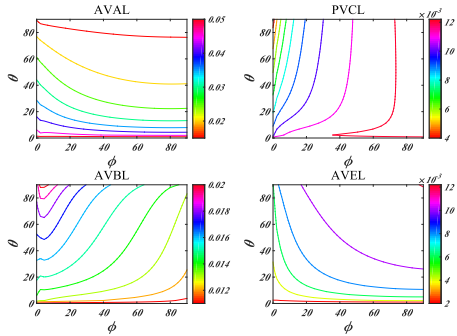}
\end{center}
\caption{(Color online) Functional Multiplex PageRank of the neurons AVAL, AVBL, PVCL and AVEL in the duplex brain network of the nematode C. elegans.}
\label{figurefmp}
\end{figure}
{According to the  Absolute  Multiplex PageRank  the top ten central neurons are the ones displayed in  Table $\ref{Tablece}$. In the top four positions of the rank list we found the AVA interneurons and the AVB interneurons. 

The Pearson correlation coefficient analysis performed on the top ten ranked node (Figure \ref{figurerho}) shows, as expected, that neurons of the same type (AVA,AVB,PVC,AVE) have  similar interaction channels, which is due to their characteristic role and functions in the nervous system.   
Also the AVDR and  the DVA are much more correlated between each other and with the AVE neurons than with the other neuron types (AVA,AVB,PVC). 

In Figure \ref{figurefmp} we show the Functional Multiplex PageRank (pattern to success) of different types of neurons in the parameter space $(\phi,\theta)$.  Given this duplex structure, the Functional Multiplex PageRank calculated  exclusively  using electric junctions  corresponds to the value at $\phi=0(0^o),\theta=\pi/2 (90^o)$ the one calculated using exclusively synaptic connections corresponds to the value  at $\phi=\pi/2 (90^o),\theta=\pi/2(90^o)$. Finally, for  $\theta=0(0^o)$,  only connections involving  both electric and chemical interactions are affecting the value of the Functional Multiplex PageRank. The  patterns to success shown in Figure \ref{figurefmp} reveal  the very different functional roles of the corresponding neurons in the brain network.
}        
\begin{figure}
\begin{center}
\includegraphics[width=0.8\columnwidth]{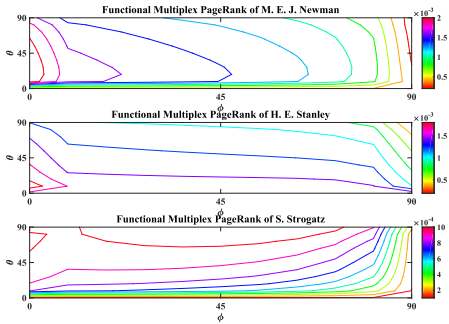}
\end{center}
\caption{(Color online) Functional Multiplex PageRank of the three important authors of PRE,  M.E. J. Newman, H. E. Stanley and S. Strogatz. The correlation $\rho$ between their Functional Multiplex PageRank is $\rho=0.504$ (Newman/Stanley), $\rho=0.780$ (Newman/Strogatz) and $\rho=0.003$ (Stanley/Strogatz). This calculation is performed on a grid with $\delta \theta=\delta\phi=\pi/20$.}
\label{figure2}
\end{figure}
 
As a third example of duplex network we consider the dataset studied in \cite{Menichetti} formed by the authors of PRE until the year 2009. The first layer is the (directed) citation network between the authors in the journal PRE, the second layer is the (undirected) collaboration network between the authors in the same journal. Both layers are here unweighted.
The analysis of the functional PageRanks of PRE authors reveals different patterns to  success, which eventually combine the centrality in the citation and the collaboration network.
Here the angle $\phi$ will tune the influence of the citation with respect to the collaboration network, while the angle $\theta$ tunes the relevance of the multilinks $(1,1)$ formed by authors that collaborate and cite other authors.
In Figure $\ref{figure2}$ we display the Functional Multiplex PageRank of three major authors,  M. E. J. Newman, H.E. Stanley and S. Strogatz showing different patterns to success.

\section{Modelling the dependence on the multiplicity of the overlap and applications to real datasets}

In this section we will study the application of the Functional Multiplex PageRank to multiplex networks with arbitrary number of layers.
Due to the exponential growth of the  number of multilinks with the number of layers $M$, we consider the case in which the value of the influence parameter $z^{\vec{m}}$ only depends on the multiplicity of the overlap $\nu(\vec{m})$.
Specifically we take
\bea
z^{\vec{m}}=q^{\nu(\vec{m})-1},
\eea
with $q>0$.
For example for three layers we will have 
\bea
z^{(1,0,0)}=z^{(0,1,0)}=z^{(0,0,1)}=1,\nonumber \\
z^{(1,1,0)}=z^{(0,1,1)}=z^{(1,0,1)}=q,\nonumber \\
z^{(1,1,1)}=q^2.
\eea
Therefore if $q>1$ the multinks with high multiplicity of overlap will have higher influence, while for 
$q<1$ their influence will be suppressed.
 
\begin{figure}
\begin{center}
\includegraphics[width=.7\columnwidth]{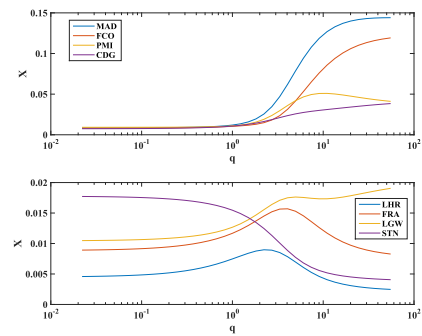}
\end{center}
\caption{(Color online) Functional Multiplex Centrality as a function of $q$ in the European airport multiplex network with $N=450$ nodes and $M=85$ layers. The airports shown are Madrid-Barajas Airport (MAD), Roma Fiumicino Airport (FCO),
Palma de Mallorca Airport (PMI), Paris Charles de Gaulle Airport (CDG) (top panel) and Heathrow Airport (LHR), Frankfurt Airport (FRA), Gatwick airport (LGW) and Stansted airport (STN) (bottom panel).}
\label{figure3}
\end{figure}

With this convention the Functional Multiplex PageRank will be a function of $q\in \mathbb{R}^{+}$.
Here we consider as an example of an application the airport multiplex network of Ref.\cite{Cardillo}, including all airline companies operating in Europe. This multiplex network is formed by $N=450$ airports connected by $N=85$ airlines.
The Absolute Multiplex PageRank rewards the major tourist destinations (see Table \ref{Table3}) that have all Functional Multiplex PageRank with a maximum for  large $q$, i.e. they are central when the influence of multilinks with large multiplicity of overlap is significant.
Nevertheless other important airports can have significantly different pattern  to success (see Figure $\ref{figure3}$).
Interestingly Heathrow and Frankfurt display a correlated pattern to success while Stansted and Gatwick airport have an anticorrelated pattern to success.
By proceeding similarly to the case of the duplex networks we can calculate the Pearson correlation which is given by $\rho=0.7641$ between Heathrow and Frankfurt airports and by $\rho=-0.9750$ between Stansted and Gatwick airports (calculated on a logarithmically spaced one dimensional grid with $q_r=e^{(r-20)/5}$ and $r=\{1,2,\ldots,40\}$).
  
\begin{table}
\caption{Top ranked airports according to the Absolute Functional Multiplex PageRank, in the  multiplex airport network formed by all $N=85$ airlines companies operating in Europe.}
\label{Table3}
\begin{center}
\begin{tabular}{cc}
Rank  & Airport \\
 1 &  Madrid-Barajas Airport (MAD)\\
 2 & Roma Fiumicino Airport (FCO) \\
 3 & Palma de Mallorca Airport (PMI)\\
 4 & Paris Charles de Gaulle Airport (CDG) \\ 
\end{tabular}
\end{center}
\end{table}

\section{Conclusions}
In conclusion we have proposed here to study the Functional Multiplex PageRank for characterizing the centrality of nodes in multiplex networks.
{This measure associates to a node a function called its {\it pattern to success} that is  able to capture the role of the different  type of connections in determining the node  centrality.}
Two generic nodes of a multiplex network can have distinct patterns leading to   their success, and here we propose a way to characterize their correlations.
From this measure we can extract an Absolute  Multiplex PageRank which provides an absolute rank between the node of the multiplex. 
We have applied this measure to airport multiplex networks,   to the multiplex connectome  of the neumatode C. elegans, and to the citation/collaboration network of PRE authors.
The Functional Multiplex PageRank can be efficiently measured on duplex multiplex networks, and when suitably  simplified, it can be applied to multiplex networks with arbitrary number of layers $M$. Interestingly this algorithm can be  easily generalized to weighted multiplex networks.

\section{Acknowledgements}
A.A. acknowledges financial support from ICREA Academia, James S. McDonnell Foundation and Spanish MINECO FIS2015-71582.

\end{document}